\documentclass[screen, manuscript,nonacm]{acmart}

\usepackage{makecell}

\newcommand{\etal}{et~al.~} 
\newcommand{\ie}{i.e.,~}
\newcommand{\eg}{e.g.,~}
\newcommand{\prototypeNameWithoutSpace}{ScrollTest}
\newcommand{\prototypeNameWithSpace}{ScrollTest~}

\AtBeginDocument{%
  \providecommand\BibTeX{{%
    \normalfont B\kern-0.5em{\scshape i\kern-0.25em b}\kern-0.8em\TeX}}}

\setcopyright{none}
\copyrightyear{2023}

\acmBooktitle{}

\begin{document}
\title{\prototypeNameWithoutSpace: Evaluating Scrolling Speed and Accuracy}

\author{Chaoran Chen}
\email{cchen25@nd.edu}
\affiliation{%
  \institution{University of Notre Dame}
  \city{Notre Dame, IN}
  \country{USA}
}

\author{Brad A. Myers}
\email{bam@cs.cmu.edu}
\affiliation{%
  \institution{Carnegie Mellon University}
  \city{Pittsburgh, PA}
  \country{USA}
}

\author{Cem Ergin}
\email{cemergin.xyz@gmail.com}
\affiliation{%
  \institution{Revel}
  \city{New York, NY}
  \country{USA}
}

\author{Emily Porat}
\email{emporat@gmail.com}
\affiliation{%
  \institution{Orchard}
  \city{New York, NY}
  \country{USA}
}

\author{Sijia Li}
\email{sijia.li.job@gmail.com}
\affiliation{%
  \institution{Plaid}
  \city{San Francisco, CA}
  \country{USA}
}

\author{Chun Wang}
\email{chun1221dd@gmail.com}
\affiliation{%
  \institution{Databento}
  \city{Salt Lake City, UT}
  \country{USA}
}

\renewcommand{\shortauthors}{Chen and Myers, et al.}

\begin{abstract}
  Scrolling is an essential interaction technique enabling users to display previously off-screen content. Existing evaluation models for scrolling are often entangled with the selection of content, \eg when scrolling on the phone for reading. Furthermore, some evaluation models overlook whether the user knows the target position. 
  We have developed \prototypeNameWithoutSpace, a general-purpose evaluation tool for scrolling speed and accuracy that avoids the need for selection. We tested it across four dimensions: 11 different scrolling techniques/devices, 5 frame heights, 13 scrolling distances, and 2 scrolling conditions (\ie with or without knowing the target position). The results show that flicking and two-finger scrolling are the fastest; flicking is also relatively precise for scrolling to targets already onscreen, but pressing arrow buttons on the scrollbar is the most accurate for scrolling to nearby targets. Mathematical models of scrolling are highly linear when the target position is unknown but like Fitts' law when known.
\end{abstract}



\begin{teaserfigure}
  \includegraphics[width=\textwidth]{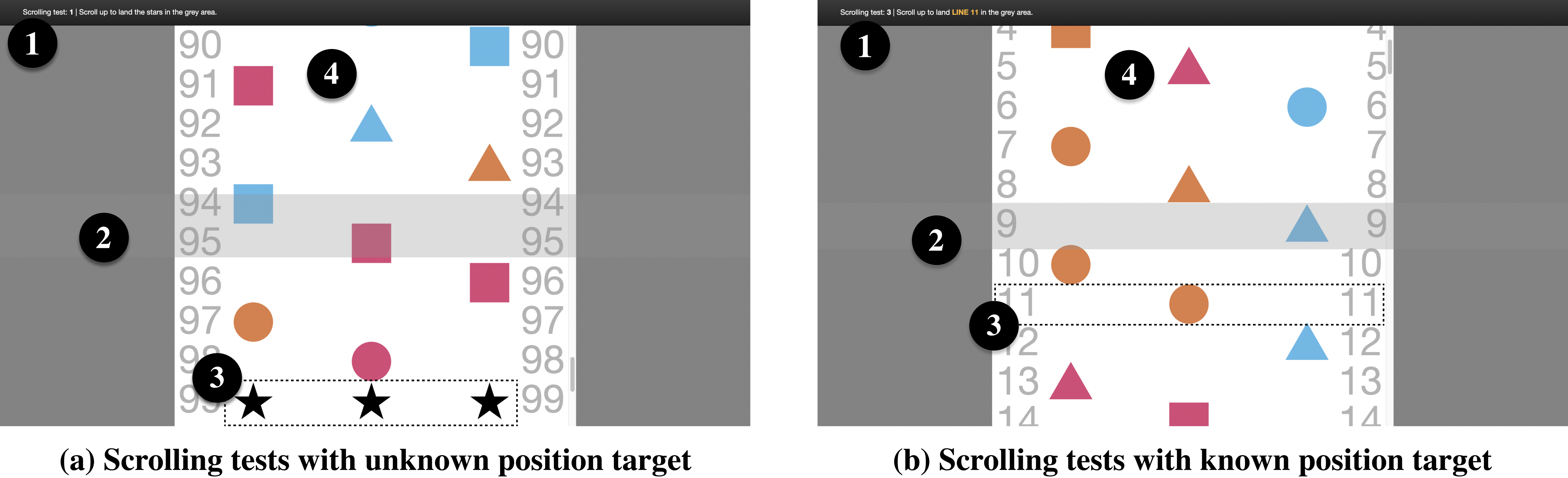}
  \caption{The user interface of \prototypeNameWithSpace contains two main pages: (a) Scrolling Tests with \textit{Unknown} Position Target and (b) Scrolling Tests with \textit{Known} Position Target. Both pages have (1) Instruction Panel, (2) Frame where the target line should be scrolled to, (3) Target Line, (4) Scrollable Area.}
  \Description{The user interface of \prototypeNameWithSpace contains two main pages, (a) Scrolling Tests with Unknown Position Target and (b) Scrolling Tests with Known Position Target. Both pages have (1)Instruction Panel, (2)Frame, (3)Target Line, and (4)Scrollable Area.}
  \label{fig:teaser}
\end{teaserfigure}

\maketitle

\section{Introduction}

\begin{figure}
    \centering
    \includegraphics[width=\linewidth]{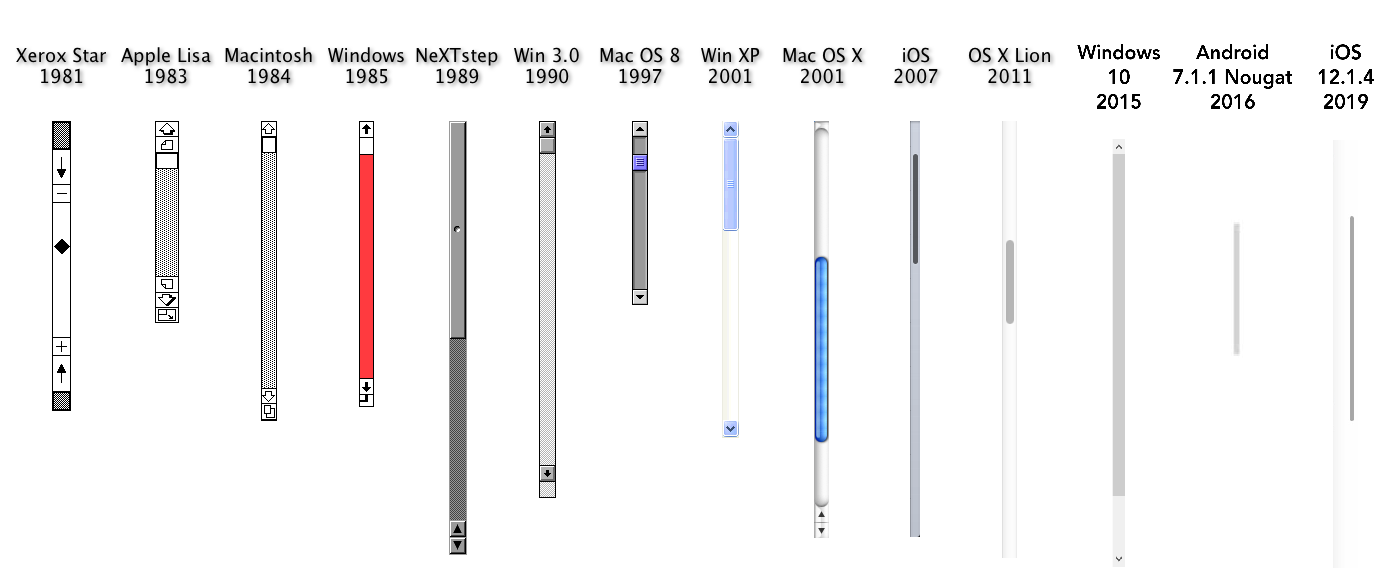}
    \caption{Some examples from the evolution of scrollbars}
    \label{fig:scrollbars}
    \vspace{-0.4cm}
\end{figure}

Scrolling is essential for people to control which part of content is visible. It has existed for over 4500 years since the original scrolls of parchment, and since at least 1974 in graphical user interfaces when scrollbars were used in the Bravo text editor and Smalltalk systems from Xerox PARC \cite{brad1990video}. 
Scrolling is now ubiquitous everywhere, including in classical desktop WIMP interfaces \cite{appert2006orthozoom,igarashi2000speed,cechanowicz2009augmented}, mobile devices \cite{miyaki2009graspzoom,baglioni2011flick,kim2014content}, car display screens \cite{ng2017evaluation} and VR/AR interfaces \cite{bagher2021move}. It is also one of the interaction techniques that has changed the most on PCs -- whereas buttons, text input fields and graphical editing are not much different from those used on the original Lisa in 1983, scrollbars have changed significantly as shown in Fig. \ref{fig:scrollbars} (\eg the arrows moving around and disappearing, support for scrolling by pages coming and going, etc.) and have also been significantly replaced on laptop by swiping gestures on their touchpads. Scrolling on touch screens such as smartphones and tablets is also quite different today than on earlier touchscreens. New scrolling techniques are also still being introduced such as for augmented and virtual reality.

Despite the ubiquity of new scrolling approaches, there is lack of standardized evaluation frameworks and metrics to support quantifying the performance of different scrolling techniques. This is in contrast to evaluating the speed of pointing with Fitts' Law \cite{fitts1954information} and its tests \cite{card1978evaluation,soukoreff2004towards,grossman2005bubble,baudisch2003drag} (some of which are even international standards \cite{standard2000ergonomic,soukoreff2004towards}), and tests for typing speeds \cite{mackenzie2002kspc,wobbrock2006analyzing}. Previous empirical studies for scrolling even show conflicting results for modeling scrolling performance: Hinckley \etal \cite{hinckley2002quantitative} reported that Fitts' law modeled the performance of scrolling techniques as a \textit{logarithmic} function, while Andersen \cite{andersen2005simple} argued \textit{against} the applicability of Fitts' law and claimed a \textit{linear} model better describes the relation between movement time and scrolling distance when the scrolling target is unknown for participants. 
To resolve the conflict, Cockburn and Gutwin \cite{cockburn2009predictive} clarified the difference of scrolling for \textit{known} versus \textit{unknown} targets in a scroll-based selection task. If the user knows the position of the target, the function of movement time can be logarithmic with distance for some scrolling techniques since the user can go much faster at the beginning if there is a long distance to cover. However, if the user does \textit{not} know how far the target is, then the user must scan at a linear speed watching for the target to come into view. 

Though Cockburn and Gutwin described a theoretical and empirical prediction model  \cite{cockburn2009predictive}, their scroll-based selection task had \textit{two} interaction techniques involved: \textit{scrolling} and \textit{selection} -- the user not only needed to move to the correct place in the document, but also to click on a target there.

In our early tests for scrolling using methods described in previous articles about scrolling \cite{cockburn2009predictive,hinckley1999touch, zhao2014model, buxton1986study, myers2000two}, all of which involve both scrolling and selecting, we found it difficult to measure the actual scrolling speed on certain devices, such as touchscreens on smartphones, where tapping on targets is slow, difficult and error prone. Additionally, selection is not always necessary, especially for scrolling tasks that only involve reading, which is a common use-case for scrolling on phones and tablets. Also when scrolling while reading, people may occasionally scroll up a little (\eg a page's worth), even though the target is already visible on the screen. This behavior has not been tested by prior research.

A further disadvantage of the prior work is that none of the previous tests discuss the issue with the \textit{accuracy} of scrolling, which might result in users overshooting the target, or going to a wrong location in the document.
Clearly, a low-accuracy method will be slower to reach the target, but it is also useful, as with text entry and pointing, to have a method that can help identify to what extent the speed is impacted by low accuracy of the scrolling itself.

Therefore, we were motivated to create a new scrolling test framework that separates scrolling from selection and known from unknown target locations, and also provides a measure for scrolling accuracy in addition to the scrolling speed. We were also interested in testing today's scrolling methods like flick scrolling on smartphones and two-finger scrolling on laptop touchpads, to see how well they perform when selection is factored out.

We present \prototypeNameWithSpace as a potential standard for testing scrolling.\footnote{\prototypeNameWithSpace will be available for general use both as-is (without any changes), and as open source code if experimenters want to make modifications. The link of \prototypeNameWithSpace is https://github.com/CharlieCRChen/scrolling-test-new/.} It runs as a browser page, so it will work on any recent screen device.
In order to minimize the effect of visual scanning and missing the target when in the unknown condition, it uses large graphical objects on the page instead of words (Fig. \ref{fig:teaser}-a). For the known condition, the user is just told which numbered line to go to (Fig. \ref{fig:teaser}-b).
\prototypeNameWithSpace provides a variety of scrolling distances -- from having the target on-screen (the ``visible'' distance) to just off-screen (``short'') to all the way down the document (``long''). In order to avoid requiring selection, we detect when the contents stop moving with the correct target (Fig. \ref{fig:teaser}: 3) in the frame area (Fig. \ref{fig:teaser}: 2), which is the area the target must be moved into. In order to test whether the scrolling times follow a Fitts' law pattern, the test uses a variety of frame area sizes. 


In this work, we address the following research questions (RQs). We also summarize our answers:
\begin{itemize}
    \item \textbf{RQ1:} Can the same test for scrolling be used for all scrolling techniques across different devices and different distances, and thereby be a candidate for a standard test? The result of our experiment shows a promising \textit{yes} to this question.
    \item \textbf{RQ2:} Which scrolling technique is fastest and most accurate for very short (visible), invisible short and invisible long distances (where the target is \underline{in}visible before scrolling begins)? We find that 
    flicking on the iPhone and iPad, two-finger scrolling, and using a mouse scroll wheel are the fastest for both known and unknown conditions and have no statistical significant differences among them. They are also statistically faster than the other techniques. Flicking is quite precise for scrolling to targets already onscreen, but pressing arrow buttons in a scrollbar is the most accurate for scrolling to nearby targets.
    \item \textbf{RQ3:} Does a linear mathematical model for unknown targets and a logarithmic model for known targets, as reported in prior papers \cite{cockburn2009predictive}, apply for all scrolling devices? Our experiment validates the models, but some scrolling devices (\eg mouse wheel with notches) do not follow that model.
    \item \textbf{RQ4:} What are the coefficients for the mathematical models mapping scrolling time and distance without clicking? Through regression analysis, we report the numbers in the result section.
    \item \textbf{RQ5:} Will the frame size impact the scrolling speed in the same way as in Fitts' law test? Our results suggest that frame size is negatively correlated with movement time as described in Fitts' law.
    \item \textbf{RQ6:} What would be practical measures for the accuracy of a scrolling method? We found that the \textit{number of switchbacks}, when the user overshoots and has to scroll in the reverse direction, serves as a good indication of scrolling accuracy, but the distance of the overshoot does not.
\end{itemize}

The main contributions of this work include:
\begin{itemize}
    \item We present \prototypeNameWithoutSpace, a general evaluation tool for measuring the speed and accuracy of scrolling techniques.
    \item We present a quantitative model and metrics to evaluate the performance of scrolling alone, which avoids the effects of clicking by just requiring stopping scrolling with the target in frame area.
    \item We conducted a user study to compare the performance of 11 scrolling techniques. The results showed statistically significant differences among techniques and conditions and the accuracy of devices varies with distance. 
\end{itemize}

\section{Related Work}
This section reviews previous work about the scrolling behaviors, novel scrolling techniques, and empirical evaluation models for scrolling.

\subsection{Scrolling Behaviors}
Prior research in scrolling behaviors studied the preferred reading regions and scrolling strategies.
Scrolling actions are initiated by readers to move the text into their preferred reading region, but one study reported that individual readers differed in whether their preferred reading regions were at the top, middle, or bottom of the screen \cite{sharmin2013reading}. Buscher \etal \cite{buscher2010eye} also analyzed the general distribution of readers' preferred reading regions and found that the visual attention was concentrated around the middle of the screen but had high variations. Sharmin \etal \cite{sharmin2013reading} validated the findings of preferred reading regions from Buscher, and found that readers' preferred reading region would get shorter as the font size increased. Turner \etal \cite{turner2015understanding} extended the two studies and took both both preferred reading regions and document structure into account. They found that scrolling generally happened at the intersection between paragraphs. Our work is informed by the above work in the setting of the position and size of the frame, which we put in the middle of the screen, with a variety of sizes.

For scrolling strategies, Hornbaek and Frøkjær \cite{hornbaek2003reading} observed that readers used non-linear navigation \cite{kim2016sats} to achieve active reading. Readers preferred skipping large blocks of text and then scrolling back to read previous text. Such a navigation method becomes more common with electronic media and suggests the need for more studies on the relationship between scrolling and reading performance \cite{brady2018scrolling}. For example, StrategicReading \cite{guo2020strategicreading} separates scrolling lengths into 3 categories: line-wise, section-wise, and page-wise. They observed that a user with better reading performance has less line-wise scrolls and more section and page-wise scrolls. In our work, we take different scrolling lengths into consideration, including visible, invisible short, and invisible long distances, and we provide results relevant to reading tasks where selection is not required. 

\subsection{Studies on Scrolling Techniques}
There is extensive research on scrolling techniques including indirect input, direct input, and edge-scrolling \cite{aceituno2017design}. Indirect input devices for scrolling, where the user touches something other than the screen, have been widely studied, including using a keyboard, touchpad, trackpoint, and mouse \cite{zhai1997dual, mcloone2005award, arthur2008evaluating}. For example,
Buxton and Myers \cite{buxton1986study} studied two-handed scrolling and selection tasks through a 1D slider on a touchpad and a mouse-like pointer. Myers \etal followed up with a study of using a Palm device in one hand to do scrolling (among other tasks) and using a mouse in the other hand to do pointing \cite{myers2000two}. 
Zhai \etal \cite{ zhai1997improving} compared scrolling web pages for targets at unknown locations using four devices: regular mouse scrollbars, regular mouse wheel, mouse with in-keyboard trackpoint, and mouse with trackpoint.
Hinckley \cite{hinckley1999touch} proposed the Scrolling TouchMouse, which enables users to issue Page Up and Page Down requests by tapping on its touch sensor. 
Bial \etal \cite{bial2010study} extended the prior work to measure 2D scrolling while using the similar devices. 

Direct input for scrolling, where the user touches screen with a stylus or finger, has received more attention recently. It allows users to scroll by flicking/dragging the interface directly \cite{aliakseyeu2008multi} or by manipulating a scrollbar-like widget on the screen, which is more intuitive and natural. For example, MultiScroll \cite{fitchett2010multiscroll} was invented to combine two complementary scrolling techniques on mobile devices: rate based scrolling (\eg using the Trackpoint in center of the Lenovo Thinkpad keyboard -- Fig. \ref{fig:apparatus} (h)) and zero order scrolling (\eg scrolling by dragging the content on touchscreen). Quinn \etal \cite{quinn2013touch} examined how users express scrolling intentions through flicking and reverse-engineered the transfer functions of touch scrolling systems.

Edge-scrolling or auto-scrolling is another important scrolling technique, where the content automatically starts scrolling when the cursor or finger nears the edge of the active area, often while dragging an object \cite{bardon1997method, aceituno2017design}. It enables users to reach out-of-view targets while extending the selection or while performing a drag-and-drop. We do not involve this category in our tests as it requires users to first select a certain text or graphic, which entangles scrolling with selection.

In our work, we studied eleven scrolling techniques across four devices, covering both indirect and direct scrolling techniques. We did not include some scrolling techniques that are rarely used in our daily life (\eg gaze-controlled scrolling \cite{turner2015understanding} and tilt-based scrolling \cite{liu2019tilt}), but \prototypeNameWithSpace could be easily applied to these.

\subsection{Empirical Evaluation Models for Scrolling}
Previous studies show conflicts in modelling scrolling performance. Hinckley \etal \cite{hinckley2002quantitative} used a reciprocal scrolling and tapping task, where users repeatedly scrolled to two known targets in a document, to show that scroll-based target acquisition is subject to Fitts’ Law. Zhao \etal \cite{zhao2014model} extended the bidirectional scrolling task into touch-sensitive displays. They also applied a logarithmic function to model the scrolling time and moving distance.
However, Andersen \cite{andersen2005simple} doubted the applicability of Fitts' law in scrolling when users did \textit{not} know the position of target before scrolling. Instead, empirical data of movement time showed a strong linear fit with scroll distance. To resolve the contention, Cockburn and Gutwin \cite{cockburn2009predictive} clarified the influence of knowing target position on the scrolling performance during a navigation and selection task. If the user knows the position of the target ahead of time, the function of movement time is subject to Fitts' law, but if not, then the function is linear. Most of the aforementioned work entangles selection with scrolling, which may involve confounding factors into the analysis. Our work tests a wider range of scrolling techniques and avoids requiring clicking.

For the measurement of scrolling accuracy, previous work had used erroneous clicking to measure the accuracy of the combined scrolling and clicking tasks \cite{hinckley2002quantitative}, when accuracy was even considered. This is not applicable to our work since we completely remove clicking from the scrolling test. In Multi-flick \cite{aliakseyeu2008multi}, the authors propose the ``number of crossings'' as a way to measure scrolling performance. It measures the number of times that the target crosses the frame area. But this metric does not take into account all of the ways that switchbacks can happen  (\eg people might scroll back and forth even if the target does not cross the frame region). Our work builds on this accuracy metric to use the number of switchbacks and maximum overshoot distance and tests them in the user study.

\section{Design Motivation and Pilot Study}
\prototypeNameWithSpace derives from a homework assignment in an HCI class in 2019, which required students to test scrolling to a known position using the same task as Buxton and Myers \cite{buxton1986study}, and to an unknown place using the same task as Zhai \etal \cite{zhai1997improving}, but people kept not seeing the target in the task, and tapping on the target was too hard on smartphones in both tasks. Therefore, a class project by some of this paper's authors investigated creating a new tool to evaluate scrolling. 

We identified five problems of existing scrolling tests that motivated our design:
1) The placement of the instructions distracted from the scrolling experience.
2) The testing environment was not standardized across all tests, resulting in unreliable results.
3) The user interface was outdated and did not accurately represent modern scrolling scenarios.
4) Some critical metrics were missed, such as the scrolling distance (long vs. short) and overshoot.
5) The old test required the user to point and click on the target after scrolling, which skewed the results.
In light of these issues, we converted the textual content into shapes with an accessible color palette for easier visual searching. The three columns take up the entire width of the scrolling area, which minimizes distraction in the scrollable area. 

We implemented the new design and used it for a homework assignment in an HCI class in spring of 2022 with 22 students. The biggest issue with their results came from the variability of their testing devices. Students got very different results from using similar but not identical scrolling devices for different participants, like touchpads on laptops of all different sizes and resolutions and speed settings in different operating systems. Moreover, students reported that clicking the start buttons in that design required additional movement of the pointer. They also suggested considering more metrics for measuring the scrolling accuracy.

Based on the results of the in-class use, we made the following adjustments: 1) We made the start button a large and easy-to-hit popup in the center of the scrollable area to minimize the effects of moving the pointer. 2) To come up with other potential measurements for accuracy, we observed that people tend to scroll back and forth when they find it hard to scroll to the desired area. Thus, we decomposed this behavior into two metrics: the number of switchbacks and the maximum overshoot distance. We took both of them into account as measures of accuracy (see next section). The reason for not using the sum of the total overshoot distances is that it confounds the number of switchbacks with the maximum distance, since the sum of the overshoot distances will increase when either the number of switchback or the maximum overshoot distance increases, whereas the maximum distance is an independent measure. 3) We moved the frame (that participants need to move the target line into) to be in the vertical \textit{center} of the screen rather than the top, and provide for different frame heights, which makes our results comparable to previous work \cite{hinckley2002quantitative, andersen2005simple} and also to follow many people's reading preferences \cite{buscher2010eye}.

\begin{figure}
    \centering
    \includegraphics[width=\linewidth]{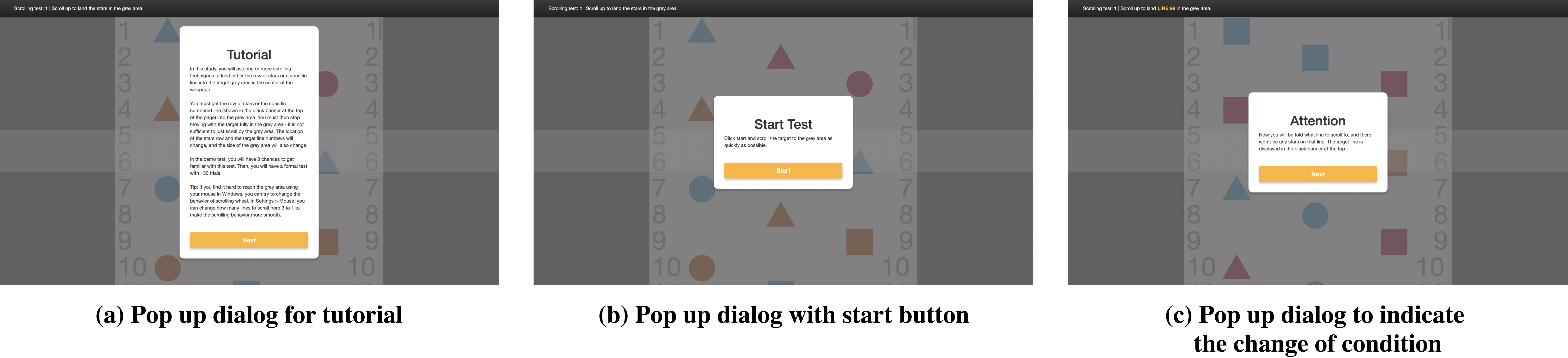}
    \caption{Design of pop up windows in \prototypeNameWithoutSpace}
    \label{fig:pop up windows}
    \vspace{-0.6cm}
\end{figure}

\section{Final System Design}
The user interface of \prototypeNameWithSpace contains two main pages: scrolling tests with \textit{\textbf{un}known} position target (Fig. \ref{fig:teaser}-a) and with \textit{known} position target (Fig. \ref{fig:teaser}-b). 

In both pages, \prototypeNameWithSpace provides instruction panels (Fig. \ref{fig:teaser} (a):1, (b):1). For targets with unknown positions, the instruction is always ``Scroll up to land the stars in the grey area'' and the target line contains three black stars (Fig. \ref{fig:teaser} (a):3). For the known condition, the instruction panel tells the participant the index of target line, as in ``Scroll up to land LINE 11 in the grey area'' (Fig. \ref{fig:teaser} (b):3). 
The size of each row of shapes is equal to three conventional lines of text, because three lines of text is the default setting in Windows 10 for how far the page moves per increment. Additionally, larger targets with line numbers make it easier for users to find the right line in the test.
The frame (Fig. \ref{fig:teaser} (a):2, (b):2) is in the middle of the screen with 5 options for its size, indicating where the target region needs to be scrolled into. 
Participants can only scroll within the scrollable area (Fig. \ref{fig:teaser} (a):4, (b):4). We intentionally set the scrollable region to be a square with ten lines in the center of the screen. It can automatically fit into the screen based on the width and height of the browser window, so no matter what size of device we test, we can insure that people can see all the ten lines of shapes (see the scrolling area on a laptop, iPhone and iPad screens in Fig. \ref{fig:apparatus}).

In the backend, \prototypeNameWithSpace records the movement time, number of switchbacks due to overshoots, and the maximum overshoot distance for each trial. We use the movement time to measure scrolling speed and the number of switchbacks and the maximum overshoot distance to measure accuracy. The timing starts right after the participant clicks the pop-up start button in the center (Fig. \ref{fig:pop up windows} (b)).

\section{Implementation Details}

We implemented \prototypeNameWithSpace with straight JavaScript and JQuery, to maximize its portability, and so we would not need to rely on any heavy extra frameworks. The database that stores the results is built using Google Spreadsheets and a Google App Script. This makes it easy to collect and analyze the results, and experimenters can switch to a different database by creating their own new Google Sheet and slightly modifying the App Script. 

In order to avoid requiring a click in the target, \prototypeNameWithSpace must detect when the target line has stopped moving, and whether it is inside the current size of the frame area. It keeps track of the position of the target via the "scrollTop" method in JQuery and compares it with the position of the frame region. We require that the target is entirely inside the frame region --- rather than merely intersecting it. To detect whether the target has stopped moving, we use a timeout of 66 milliseconds and make sure that there are no scrolling events for that amount of time. This number was empirically determined through many iterations to make sure that it is not triggered by accident when the user did not intend to stop, but is not too long that the user thinks the system did not notice it had stopped. We detect the first overshoot by watching for the scrollTop to go past the top of the frame area, and then count the rest and measure the maximum overshoot by watching for changes in the direction of the scroll movement.

In earlier versions of \prototypeNameWithoutSpace, there were problems with making sure that the keyboard focus and the mouse cursor were both in the scroll area so that keyboard arrow keys and mouse wheel or laptop two-finger scroll gestures would be sent to the correct area. In addition to manually setting these in the code, moving the start test dialog to be in the center of the scrolling area (Fig. \ref{fig:pop up windows}) made sure that this was the case. Another interesting implementation challenge was avoiding race conditions between the user stopping and starting scrolling just as the test was resetting for the next trial. 

\section{User Study}
We used our new \prototypeNameWithSpace in a controlled IRB-approved user study to see how well it worked in measuring various scrolling techniques.

\subsection{Participants}
\begin{figure}
    \centering
    \includegraphics[width=\linewidth]{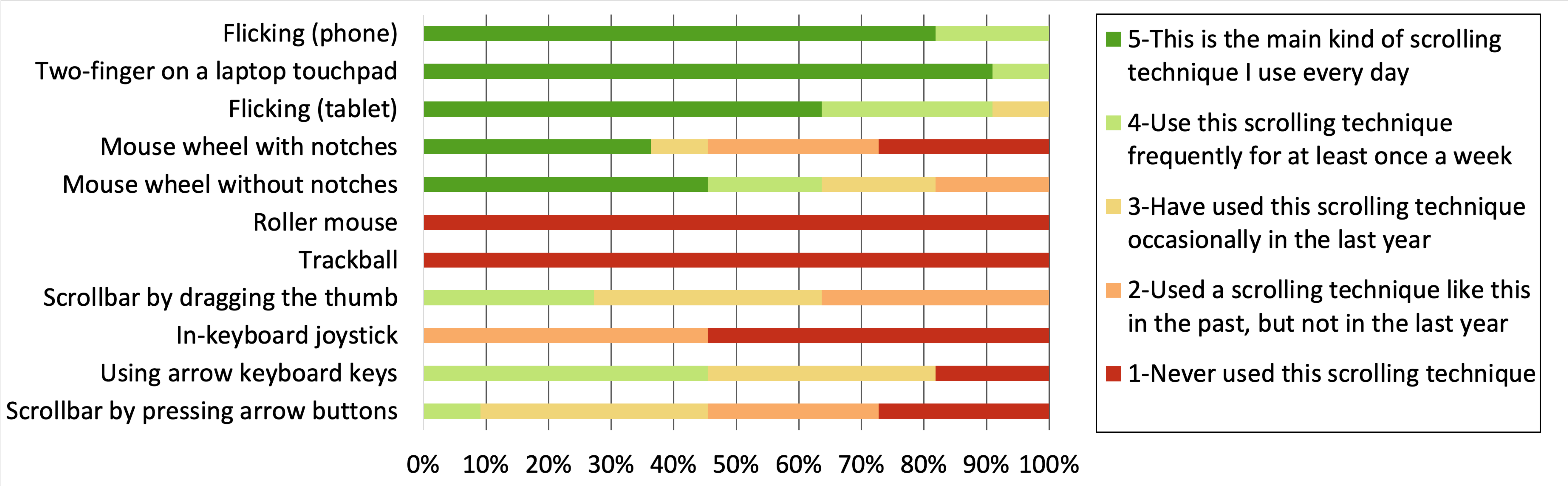}
    \caption{Participants' previous experience with the scrolling devices}
    \label{fig:participant_exp}
    \vspace{-0.4cm}
\end{figure}

\begin{figure}
    \centering
    \includegraphics[width=\linewidth]{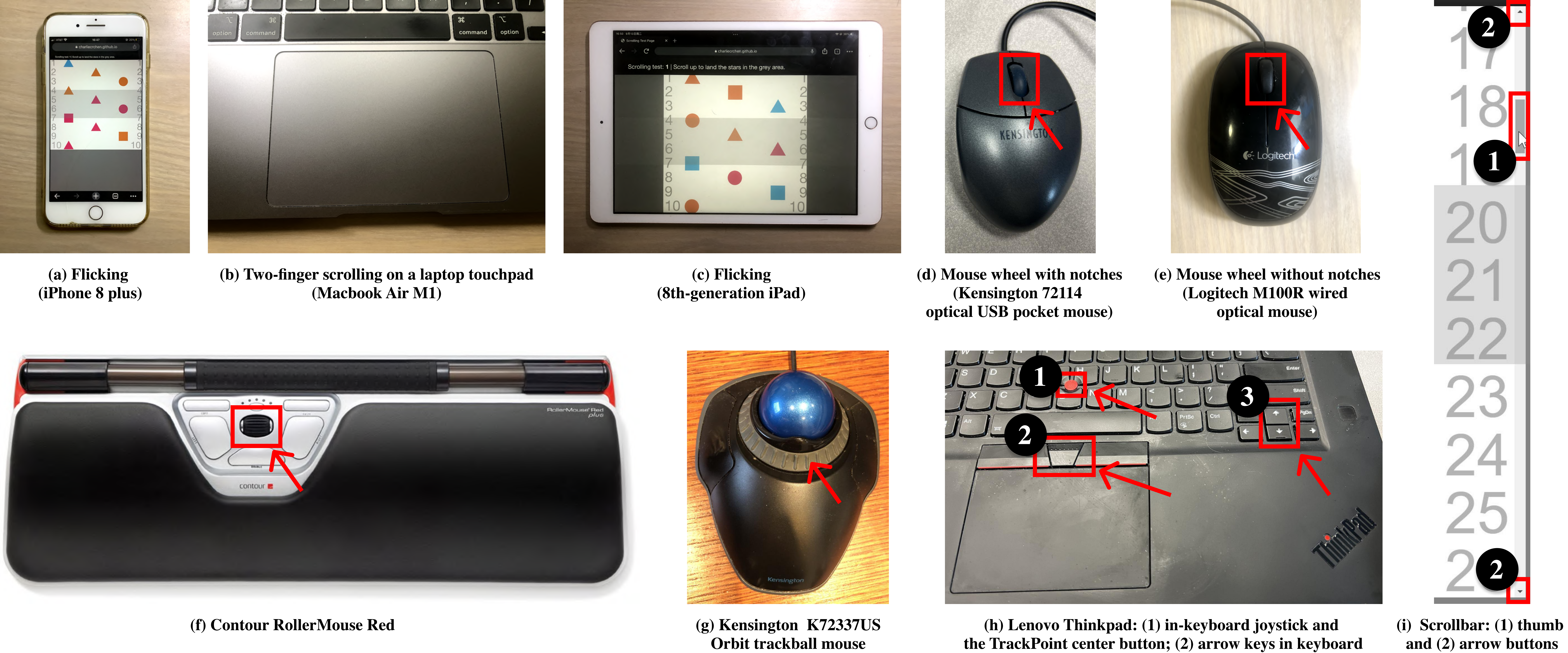}
    \caption{Scrolling devices}
    \label{fig:apparatus}
    \vspace{-0.5cm}
\end{figure}

We recruited 11 participants (9 women) with an average age of 23.64 (SD=1.77). All participants used their right hand in the test, although one participant was left-handed but usually controls the mouse cursor with the right hand. All of the participants reported that they had experience in using a laptop and/or desktop computer (proficient: 8, advanced:2, intermediate:1, Basic:0, Little or none:0), and in using a smartphone and/or tablet (proficient:9, advanced:2, intermediate:0, Basic:0, Little or none:0). Most of them considered themselves as heavy users of laptop/desktop (10 of 11) and smartphone/tablet (7 of 11), usually spending more than 5 hours per day using the devices. Half of them usually prefer using the built-in touchpad on a laptop to move the the cursor (6 of 11), while others prefer using an external mouse to move the cursor on a laptop or desktop (5 of 11). In Fig. \ref{fig:participant_exp}, we show the participants' self-reported experience of all the tested scrolling devices. Two-finger scrolling on a laptop touchpad, flicking on phones and tablets, and using a mouse wheel (with and without notches) are the most popular ways to scroll, while none of them had experience using the Contour roller mouse (Fig. \ref{fig:apparatus} (f)) or a trackball (Fig. \ref{fig:apparatus} (g)).

To limit our study to 40 minutes total, we used a hybrid between/within-subjects design where each subject used three of the eleven scrolling techniques for the study such that each device was used by 3 participants. Participants were paid \$10 for their time.

\subsection{Apparatus}
The experiment was displayed on four devices (Fig. \ref{fig:apparatus}): We used a 5.5-inch iPhone 8 plus (Fig. \ref{fig:apparatus} (a)) and a 10.2‑inch 8th-generation iPad (Fig. \ref{fig:apparatus} (c)) to test flicking on a touchscreen. We used a 13-inch MacBook Air (Fig. \ref{fig:apparatus} (b)) to test the two-finger scrolling on a laptop touchpad. For the test of other scrolling techniques, we used a 14-inch Lenovo Thinkpad running Windows 10. For two techniques, we used the Windows scrollbar as displayed by the Chrome browser (Fig. \ref{fig:apparatus} (i)).

We explain the details of scrolling techniques as follows:
\begin{itemize}
    \item \textbf{Flicking on a phone or on a tablet:} The user scrolls the contents up or down by putting a finger on the touchscreen and either dragging it up and down, or flicking it (Fig. \ref{fig:apparatus} (a)(c)). If flicked, the movement speed of the content is proportional to the speed of the flick, so users can flick faster if they want to scroll a further distance. The contents slow down and stop on their own, as if there was momentum and friction. If the user touches the content while it is moving, it stops instantly. If the user drags slowly and releases, the contents stop as well.
    \item \textbf{Two-finger scrolling on a laptop touchpad:} The user scrolls the contents by putting two fingers on the touchpad which is below the keyboard and moves up or down (Fig. \ref{fig:apparatus} (b)). The two fingers should move together to avoid zooming. The two fingers can be dragged or flicked in the same way as described for a touchscreen.
    \item \textbf{Mouse wheel with notches:} The user scrolls the contents by incrementally scrolling the mouse wheel which is at the center top of the mouse, between the two buttons (Fig. \ref{fig:apparatus} (d)). The notches cause the wheel to click with each incremental movement. The scroll wheel cannot be flicked -- it will just stop moving when the finger is lifted. We used a Kensington 72114 optical USB pocket mouse (Fig. \ref{fig:apparatus} (d)) in our test.
    \item \textbf{Mouse wheel without notches:} The mouse wheel allows user to scroll smoothly and would continue moving if flicked. We used a Logitech M100R wired optical mouse (Fig. \ref{fig:apparatus} (e)) to test this technique.
    \item \textbf{Roller mouse} by Contour: The user scrolls the contents by rolling or flicking the smooth moving scroll cylinder in the center of the panel (Fig. \ref{fig:apparatus} (f)). 
    \item \textbf{Trackball}: The user scrolls the content by rotating the grey smooth-scrolling scroll ring around the blue trackball ball (the trackball itself controls the mouse pointer position). We tested it by using a Kensington Orbit Trackball (Fig. \ref{fig:apparatus} (g)) in the experiment.
    \item \textbf{Scrollbar by dragging the thumb}: The user can scroll the content by dragging the thumb (or ``scroller'' as it is called in Mac OS) (Fig. \ref{fig:apparatus} (i)-1) with the pointer along the trough (or track). The position of the thumb in the scrollbar corresponds to the position of the visible content across the entire page. In our study, we tested it by asking user to only drag the thumb by using the touchpad in the Thinkpad as a way of scrolling the page.
    \item \textbf{In-keyboard joystick} (Trackpoint): The user can scroll the contents by holding down the TrackPoint center button  (Fig. \ref{fig:apparatus} (h)-2) and pulling or pushing the in-keyboard joystick (Fig. \ref{fig:apparatus} (h)-1), which causes the page to scroll in that direction at a rate proportional to how hard they press. This means that the user controls the direction and speed at which the document scrolls, rather than controlling the position directly. We tested this scrolling technique by using the red pointing stick (also called the IBM Trackpoint \cite{zhai1997improving}) in the Lenovo Thinkpad keyboard.
    \item \textbf{Keyboard arrow keys:} The user can move the content by pressing the up or down arrow keys on the keyboard (Fig. \ref{fig:apparatus} (h)-3). If the user holds down a key for more than about 500 milliseconds, it will start repeatedly sending the keystroke at a fixed rate of about 30 repetitions per second until the key is released.
    \item \textbf{Scrollbar by pressing arrow buttons:} The user can move the content by pressing the arrow buttons (Fig. \ref{fig:apparatus} (i)-2) at the ends of the scrollbar. If the user presses and holds down the mouse button over the arrow for more than a few milliseconds, it will also start to auto-repeat until the mouse button is released. Therefore, the user can press and hold the arrow button in the scrollbar and keep scrolling at a constant speed. We tested it by asking user to only press arrow buttons via the touchpad in the Thinkpad to scroll the targets into the frame area.
\end{itemize}

To make sure people could scroll the target into the frame, we changed the setting of the notch in the laptop from the default of 3 lines to be 1 line. All of the other mouse and scrolling speed settings in the two laptops were left at their default values.

\subsection{Task and Procedures} \label{section 6.3:Task and Procedure}
Our study design crossed the 11 scrolling techniques vs 13 different scrolling distances (D) vs 5 frame heights (H) vs. 2 conditions (known or unknown). We counterbalanced the order of devices across participants using a Latin Square.

The specific distances we used were that participants needed to scroll to lines 8, 9, 10, 11, 20, 30, 40, 50, 60, 70, 80, 90, 99. Since 10 lines were shown on the screen and each trial started with line 1 at the top, scrolling to lines 8, 9 and 10 meant the target was on the screen, so these are labeled as \textit{visible}. Scrolling to lines 11, 20, 30, 40, 50 were grouped as \textit{short}, and 60, 70, 80, 90, 99 were \textit{long}.

The specific frame heights we used were 1.0, 1.5, 2.0, 2.5, 3.0 times of the line height of the shapes.

In the user study, participants first read the online consent form. After agreeing to that, they were shown the scrolling test software, and got to practice moving up and down using their first scrolling device and technique. Then they were given a tutorial introduction to each condition of the task. They started with unknown position target, followed by the known condition test. Within each condition, they were randomly given one of the 5 frame heights, and did all 13 distances in a random order, followed by 13 distances with a different frame size, and so on.

They were optionally given a rest after doing all the trials with one device, and then were shown another scrolling technique. When they finished three devices or techniques, they answered a questionnaire about their experience and demographics.

These distances, frame heights, number of times repeated, and order of the condition are all hard-wired in the code in easy-to-edit global variables, to both make it easy to run the experiment for each participant and device, but also easy to edit if an experimenter wants to use different values. Another option, which we did not use in any user tests, requires users to click on the target line after it is in the frame area, in case future experimenters want to measure scrolling and pointing together.

\begin{table}
    \centering
    \begin{tabular}{ |c|c|c|c|c|c|c| }
         \hline
          & \multicolumn{3}{|c|}{Unknown} 
          & \multicolumn{3}{|c|}{Known} \\
         \hline
         Scrolling techniques & \makecell[c]{Time \\ (s)} & \makecell[c]{Number of \\ switchbacks} & \makecell[c]{Maximum \\ overshoot \\ distance (px)} & \makecell[c]{Time \\ (s)} & \makecell[c]{Number of \\ switchbacks} & \makecell[c]{Maximum \\ overshoot \\ distance (px)} \\
         \hline
         Flicking (iPhone)	& 2.663 & 0.856	& 326.590 & 3.384 & 0.841 & 367.251 \\
         \hline
         \makecell[c]{Two-finger scrolling on a \\ laptop touchpad}	& 3.131	& 0.964	& 270.769	& 3.362	& 1.174	& 295.374 \\
         \hline
         Flicking (iPad) & 3.207	& 0.969	& 323.713 & 3.412 & 0.953 & 290.821 \\
         \hline
         Mouse wheel with notches & 3.261 & 1.364 & 180.995	& 4.330 & 1.333 & 122.856 \\
         \hline
         Mouse wheel without notches & 3.347 & 1.897	& 163.379 & 4.258 & 1.815 & 157.395 \\
         \hline
         Roller mouse &	4.428 &	1.379 &	183.374 & 4.907	& 1.246 &	133.477 \\
         \hline
         Trackball & 4.825 & 1.554 & 96.749 & 5.432 & 2.062	& 157.226 \\
         \hline
         Scrollbar by dragging the thumb & 4.871 & 1.278	& 125.379 & 4.755 & 1.672 & 93.333 \\
         \hline
         In-keyboard joystick & 5.260 & 0.974 & 179.589	& 5.292 & 0.887 & 106.508 \\
         \hline
         Using arrow keyboard keys & 5.375 & 0.667 & 112.071 & 6.786 & 0.882 & 60.077 \\
         \hline
         Scrollbar by pressing arrow buttons & 6.443	& 0.723 & 56.713 & 7.791 & 0.748 & 81.749 \\
         \hline
    \end{tabular}
    \caption{Average time and accuracy for all scrolling devices in both conditions}
    \label{tab:time_accuracy}
    \vspace{-0.5cm}
\end{table}

\begin{table}
    \begin{tabular}{ |c|c|c|c|c|c|c|c|c|c| }
         \hline
          & \multicolumn{3}{|c|}{Unknown (linear fit)} 
          & \multicolumn{3}{|c|}{Known (linear fit)} 
          & \multicolumn{3}{|c|}{Known (log fit)} \\
         \hline
         Scrolling techniques & a & b & ${R^2}$ & a & b & ${R^2}$ & a & b & ${R^2}$ \\
         \hline
         Flicking (iPhone)	& 1.109 & 0.035 & 96.72\% & 2.179 & 0.021 & 78.00\% & -1.42 & 0.949 & 86.25\% \\
         \hline
         \makecell[c]{Two-finger scrolling on a \\ laptop touchpad} & 1.376 & 0.040 & 85.99\% & 2.431 & 0.016 & 82.42\% & 0.057 & 0.661 & 82.28\% \\
         \hline
         Flicking (iPad) & 1.256 & 0.044 & 94.16\% & 2.287 & 0.021 & 79.98\% & -1.318 & 0.949 & 90.70\% \\
         \hline
         Mouse wheel with notches & 1.631 & 0.037 & 96.43\% & 3.234 & 0.018 & 70.82\% & 0.351 & 0.783 & 72.15\% \\
         \hline
         Mouse wheel without notches & 1.333 & 0.045 & 94.13\% & 2.594 & 0.025 & 88.67\% & -0.959 & 1.008 & 79.22\% \\
         \hline
         Roller mouse & 2.252 & 0.049 & 82.63\% & 3.349 & 0.025 & 75.04\% & -1.148 & 1.178 & 82.06\% \\
         \hline
         Trackball & 2.089 & 0.062 & 93.99\% & 4.06 & 0.023 & 61.10\% & 0.002 & 1.07 & 66.62\% \\
         \hline
         Scrollbar by dragging the thumb & 2.466 & 0.054 & 93.57\% &	4.277 & 0.013 & 66.47\% & 2.191 & 0.559 & 79.86\% \\
         \hline
         In-keyboard joystick & 2.631 & 0.059 & 92.71\% & 3.646 & 0.029 & 84.51\% & -1.087 & 1.278 & 91.10\% \\
         \hline
         Using keyboard arrow keys & 2.306 & 0.069 & 97.26\% & 4.405 &	0.042 & 78.29\% & -2.416 & 1.847 & 83.60\% \\
         \hline
         Scrollbar by pressing arrow buttons & 4.033 & 0.054 & 86.42\% & 4.764 & 0.048 & 66.53\% & -4.206 & 2.317 & 74.83\% \\
         \hline
    \end{tabular}
    \caption{Coefficients of each device for the mathematical models mapping scrolling time and distance}
    \label{tab:device_coefficients}
    \vspace{-0.5cm}
\end{table}

\section{Results}

\begin{figure}
    \centering
    \includegraphics[width=\linewidth]{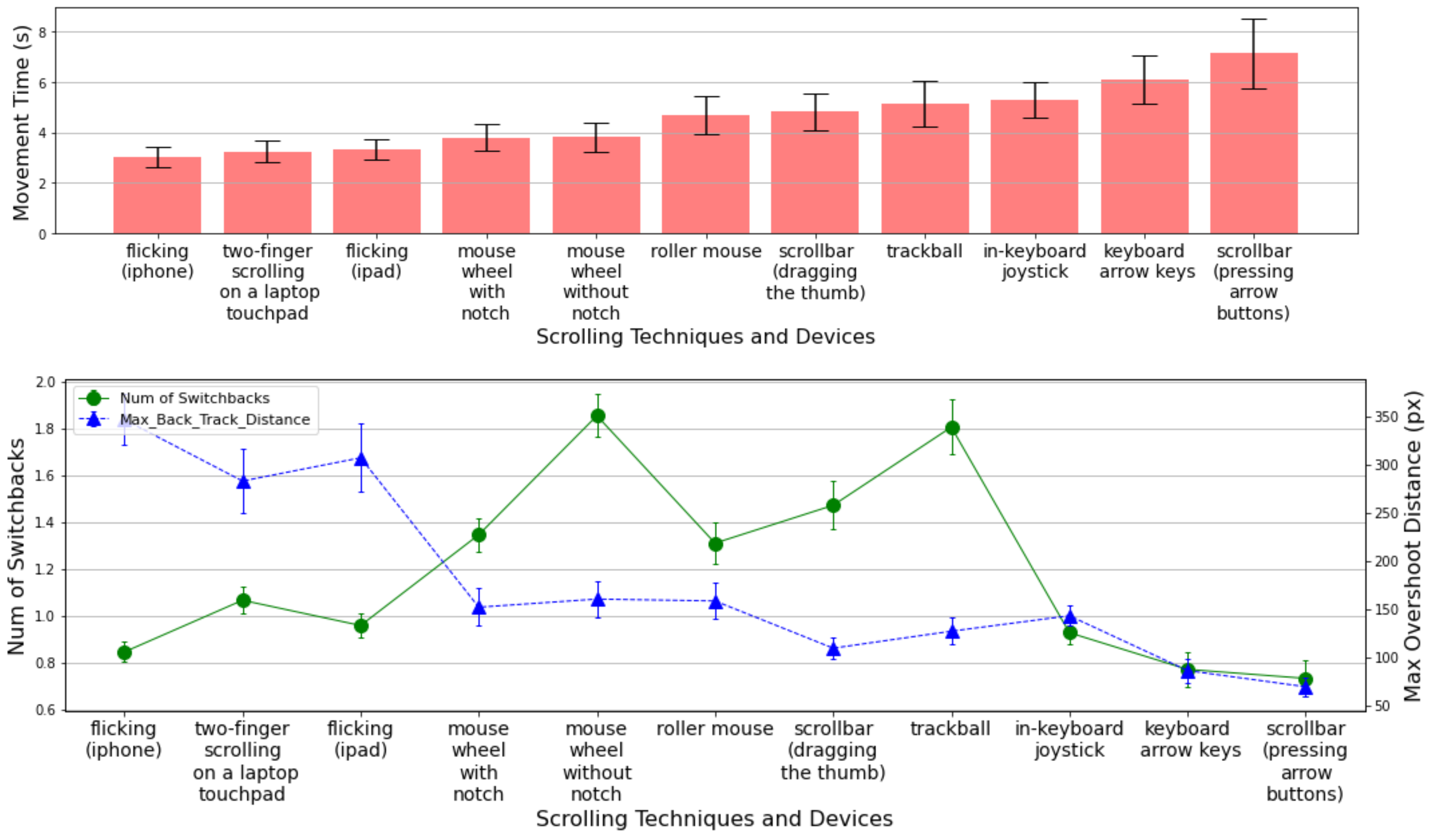}
    \caption{Average movement time (top) (shorter is better), and accuracy (bottom) (number of switchbacks and maximum overshoot distance) for all devices across all frame heights, distances, and conditions.}
    \label{fig:performance_across_devices}
\end{figure}

Fig. \ref{fig:performance_across_devices} and Table \ref{tab:time_accuracy} show the aggregated mean movement time and accuracy across all distances, frame sizes, and conditions. These provide evidence that \prototypeNameWithSpace can be used for different scrolling techniques across different devices and different distances, and thereby can be a candidate for a standard test (\textbf{RQ1} in the introduction). Flicking on mobile devices is the fastest while using scrollbar by pressing arrow buttons is the slowest but most accurate (\textbf{RQ2}). In Table \ref{tab:time_accuracy}, we show the movement time and accuracy for all devices in both unknown and known conditions. Flicking is the fastest for unknown condition while two-finger scrolling on a laptop touchpad is the fastest for known condition. For both conditions, using scrollbar by pressing arrow buttons is the slowest. We ran an ANOVA and Tukey’s Honestly Significantly Differenced (HSD) test \cite{abdi2010tukey} to measure specific differences between all device pairs. Significant differences of movement time are observed among the devices (${F_{unknown}(10,2134)=35.67}$, ${p_{unknown}<0.001}$, ${F_{known}(10,2134)=38.19}$, ${p_{known}<0.001}$). Pairwise comparison show that the 11 devices can be classified into 3 groups:
\begin{itemize}
    \item Group 1: Flicking (iPhone), Two-finger scrolling on a laptop touchpad, Flicking (iPad), Mouse wheel with notches, Mouse wheel without notches
    \item Group 2: Roller mouse, Trackball, Scrollbar by dragging the thumb, In-keyboard joystick
    \item Group 3: Using keyboard arrow keys, Scrollbar by pressing arrow buttons
\end{itemize}
For both conditions, each pair of devices within a group has no significant difference, while the pairs between groups show significant differences (${p<0.01}$). 
Table \ref{tab:correlation_time_accuracy} shows a trade-off between accuracy and speed. There is a strong positive correlation between the average movement time and the average number of switchbacks, which is expected and provides confidence in the results.

\begin{table}
    \centering
    \begin{tabular}{|c|c|c|c|}
    \hline
    Metrics 1 & Metrics 2 & ${Pearson's \ r_{unknown}}$
     & ${Pearson's \ r_{known}}$ \\
    \hline
    Movement time (s) & Number of switchbacks & 83.84\% & 83.72\% \\
    Movement time (s) & Maximum overshoot distances (px) & 18.60\% & 43.70\% \\
    Number of switchbacks & Maximum overshoot distances (px) & 59.37\% & 13.90\% \\
    \hline
    \end{tabular}
    \caption{Correlation between the average of  movement time and the accuracy for all distances across all devices and frame heights}
    \label{tab:correlation_time_accuracy}
    \vspace{-0.5cm}
\end{table}

\begin{figure}
    \centering
    \includegraphics[width=\linewidth]{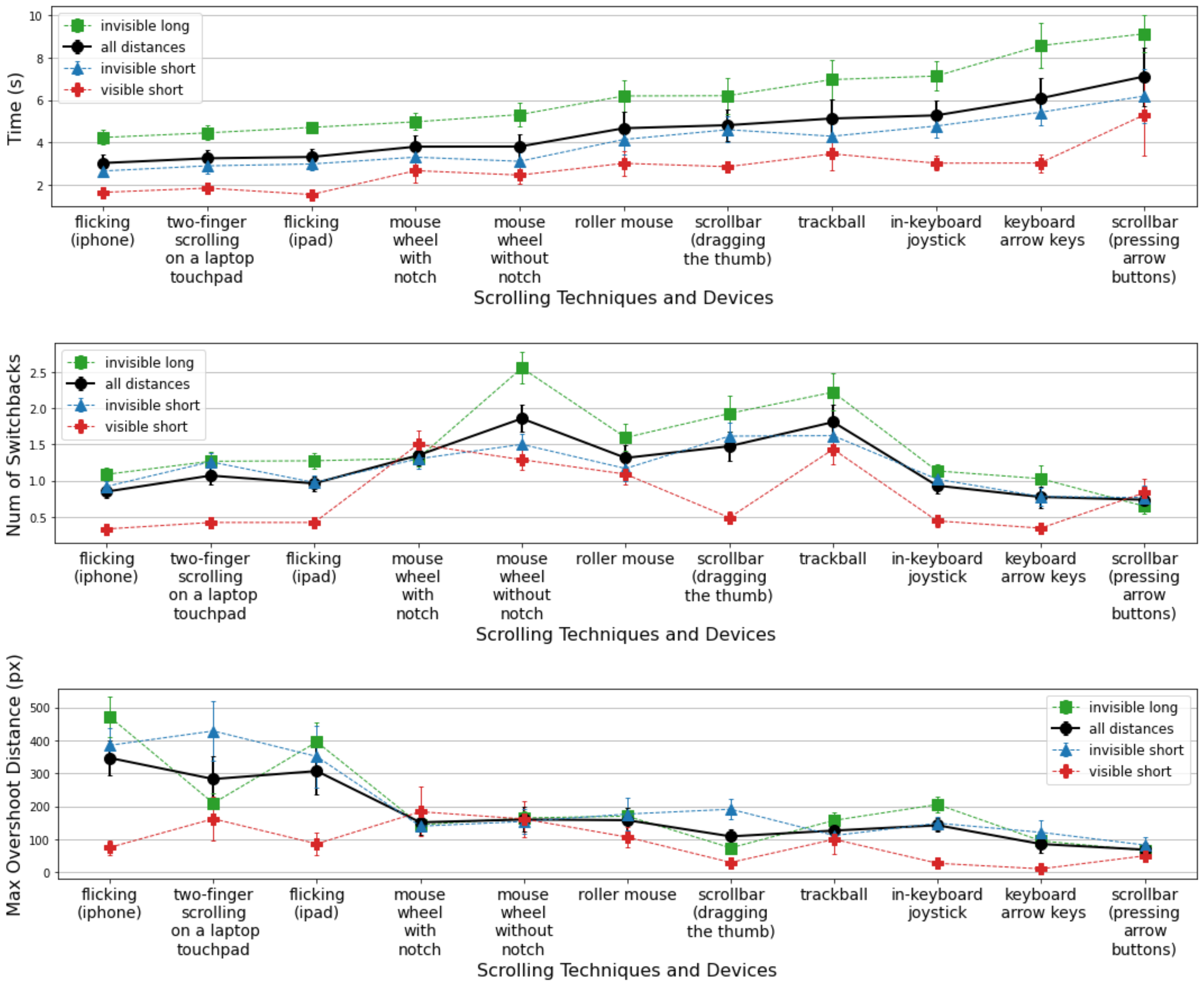}
    \caption{Mean movement time and mean accuracy (number of switchbacks and max overshoot distance) for all devices for the three distance groups}
    \label{fig:performance_across_distance}
    \vspace{-0.05cm}
\end{figure}

\begin{figure}
    \centering
    \includegraphics[width=\linewidth]{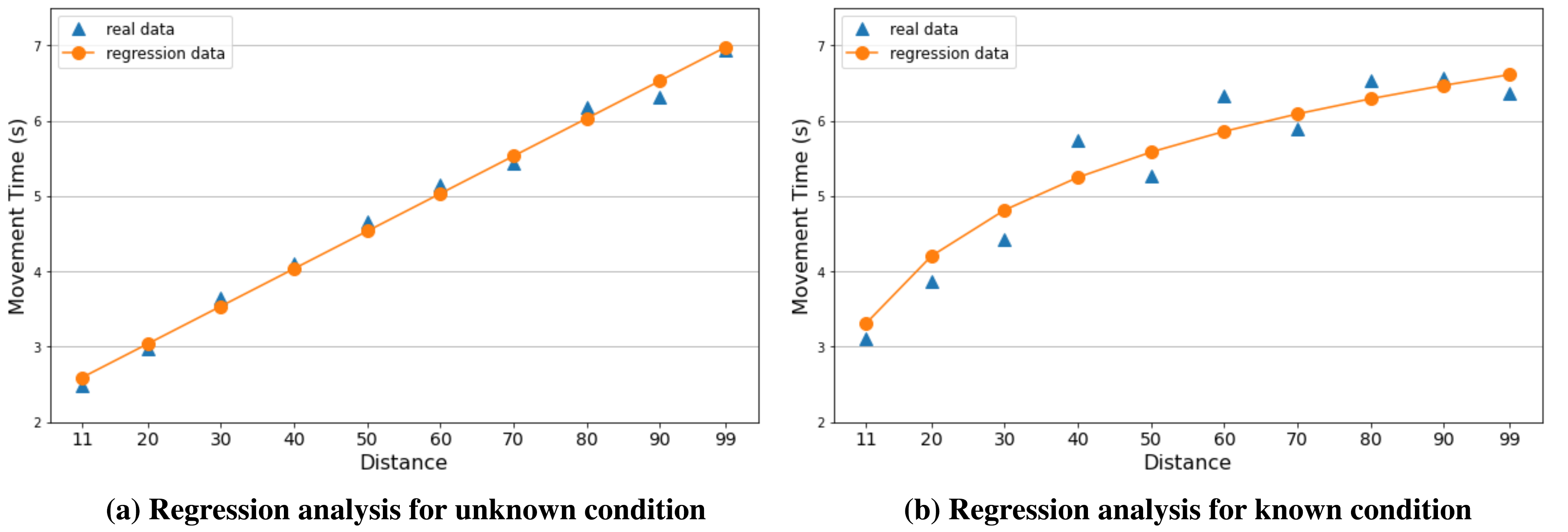}
    \caption{Regression analysis of average movement time and distance across all devices and frame sizes.}
    \label{fig:time_distance_2Conditions}
\end{figure}

We further investigate the relationship between the scrolling performance and scrolling distances. For ease of analysis, we group the distances into 3 groups (\ie visible, short, and long as mentioned above in section \ref{section 6.3:Task and Procedure}). The mean movement time (Fig. \ref{fig:performance_across_distance}-top) shows consistent results for all groups of distances. Flicking is the fastest for unknown condition and two-finger scrolling on a laptop touchpad is the fastest for known condition. Pressing arrow buttons in a scrollbar are the slowest across all distances. Flicking is also relatively precise for scrolling to targets already onscreen, but pressing arrow buttons on a scrollbar is the most accurate for scrolling to nearby targets (Fig. \ref{fig:performance_across_distance}-bottom).

We use the same way as described in \cite{cockburn2009predictive} to define the fit model of mean movement time T and scrolling distance D: a linear model ${T = a + b \times D}$ for the unknown condition and a logarithmic model ${T = a + b \times log_2(D)}$ for the known condition, where a and b are coefficients empirically obtained through regression analysis. 
Fig. \ref{fig:time_distance_2Conditions} shows the average movement time and distance for both conditions. In the unknown condition where people do not know the position of target, the movement time is linear with scrolling distances (${R^2}$ = 0.993). In the known condition where people know the position of target, the movement time is logarithmic with distance (${R^2}$ = 0.932). The overall results validate the results reported in previous work \cite{cockburn2009predictive}, \ie the linear function for unknown targets and logarithmic for known targets (\textbf{RQ3}).
The coefficients for the mathematical models (\textbf{RQ4}) are as follows:
\[
Unknown, R^2=0.993, T = 2.044 + 0.05 \times D
\]
\[
Known, R^2=0.932, T = -0.302 + 1.043 \times log_2(D)
\]

\begin{figure}
    \centering
    \includegraphics[width=\linewidth]{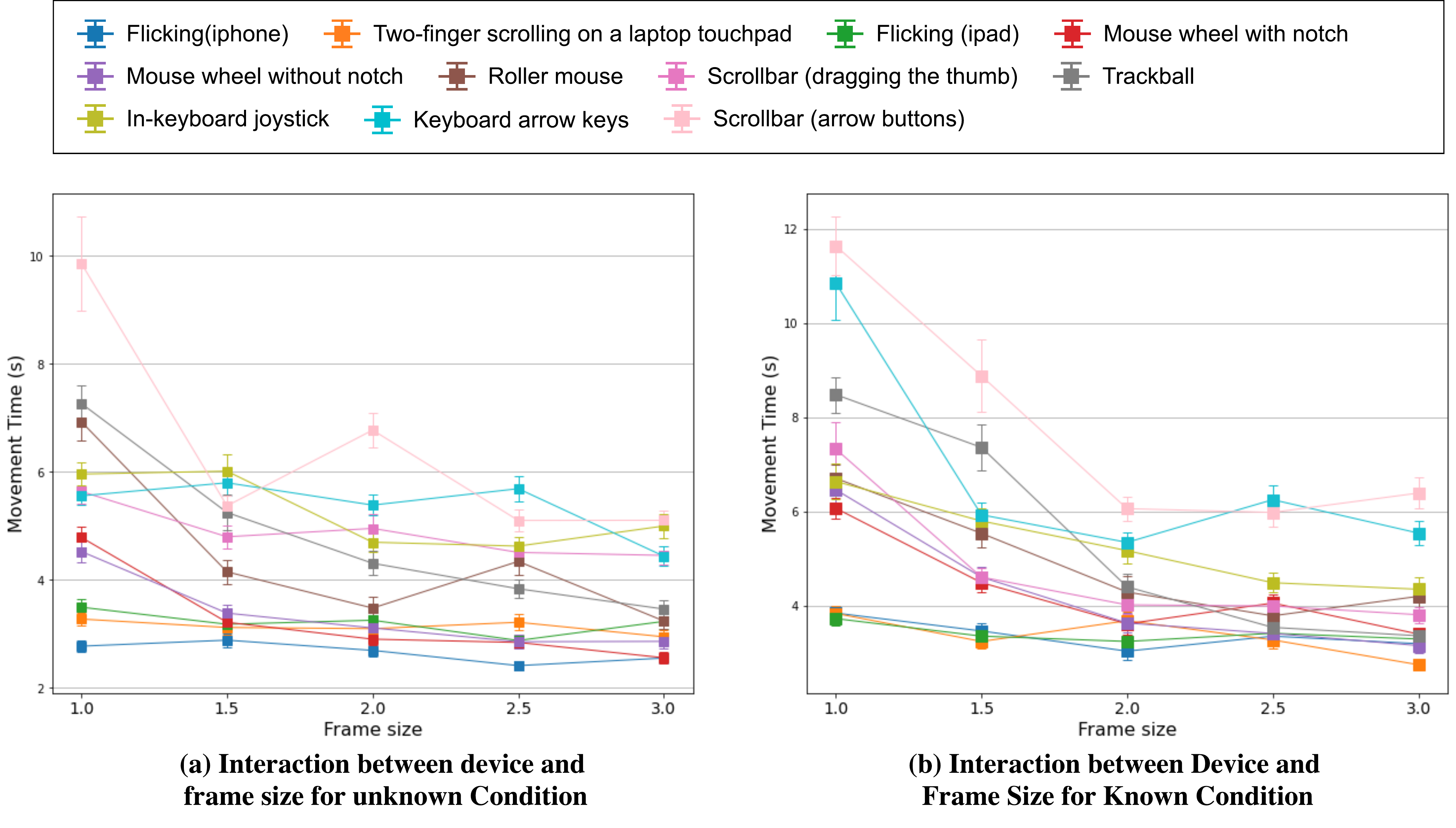}
    \caption{Interaction of mean movement time between devices and frame sizes for two conditions}
    \label{fig:time_across_frame_sizes}
    \vspace{-0.4cm}
\end{figure}

\begin{figure}
    \centering
    \includegraphics[width=\linewidth]{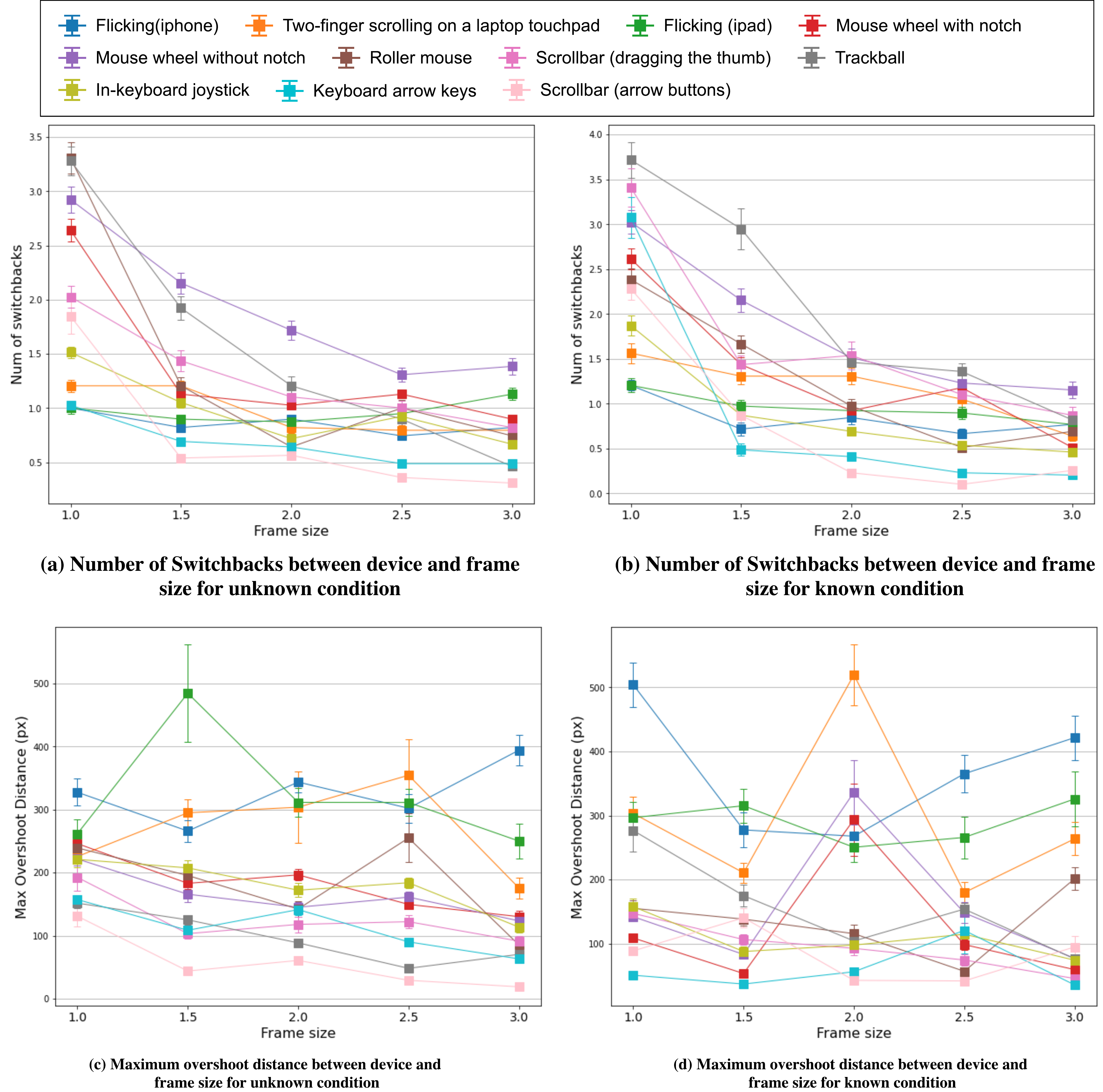}
    \caption{Interaction of accuracy between devices and frame sizes for two conditions}
    \label{fig:accuracy_across_frame_sizes}
    \vspace{-0.4cm}
\end{figure}

However, not all devices follow the models. Table \ref{tab:device_coefficients} shows the coefficients ${a}$, ${b}$ and the the coefficient of determination ${R^2}$ for all devices in both conditions. In the unknown condition, The ${R^2}$ of Roller mouse (${R^2} = 82.63\%$), Two-finger scrolling on a laptop touchpad (${R^2} = 85.99\%$), and Scrollbar by pressing arrow buttons (${R^2} = 86.42\%$) are lower than 90\%, denoting that they do not fit the linear regression model very well. As for the known condition, we compare the ${R^2}$ for both linear fit and log fit. Most devices have higher ${R^2}$ with log fit, except mouse wheel without notches.

Turning now to the frame size, our results suggest that the frame size has the same impact on movement time as described in Fitts’ law (\ie the movement time is negatively correlated with the frame size, so getting the target line into a smaller area is slower) (\textbf{RQ5}). Fig. \ref{fig:time_across_frame_sizes} shows a negative correlation (${Pearson's \  r_{unknown} = -90.73\%}$, ${ Pearson's \  r_{known} = -89.87\%}$) between the frame size and the mean movement time across the scrolling devices, with some devices being more affected than others. We also investigated the relation between frame size and the two metrics of scrolling accuracy. The number of switchbacks shows a negative correlation with the frame size (${Pearson's \ r_{unknown} = -88.09\%}$, ${Pearson's \ r_{known} = -91.41\%}$). This means people are more likely to miss the area for smaller frame sizes, as expected. But the maximum overshoot distance is insensitive to the change of frame sizes in the known condition (${Pearson's \ r_{unknown} = -93.38\%}$, ${Pearson's \ r_{known} = -0.56\%}$).

\section{Discussion}

The mean movement time in our experiment ranges from 2 seconds to 8 seconds, which is similar to the range reported in some prior work \cite{hinckley2002quantitative, andersen2005simple, cockburn2009predictive}. For example, the mean scrolling time in Andersen' study \cite{andersen2005simple} is around 6 seconds for 200 lines of text and about 8 seconds for 400 lines of text. Specifically, Andersen' study used an optical mouse with a scrollbar. The mean movement time for the scrollbar in our paper ranges from 4 to 8 seconds. Additionally, the scrolling distance in \cite{andersen2005simple} is similar to our experiment with 99 rows of shapes, because 1 row in our experiment equals 3 lines of text. 
However, there are marked differences of the scrolling time between Zhai \etal's study \cite{zhai1997improving}, where the mean movement time ranges from 50 seconds to 80 seconds. The difference may come from the different task design: Zhai's study required participant to scroll and point at a hyperlink in a long page of text. Thus, the time reported in \cite{zhai1997improving} confounds the times of scrolling, visual searching for a relatively small and inconspicuous target, and pointing, which together total much longer than our results.

The coefficient in our mathematical models ($R^2_{linear}$=0.993, $R^2_{log}$=0.932) achieve similar overall performance compared to those in \cite{cockburn2009predictive} ($R^2_{linear}$=0.99, $R^2_{log}$=0.96). However, Cockburn \etal only tested scrolling by letting participants pick whether they wanted to use a mouse wheel and/or a scrollbar. Our tests generalize their conclusion to show it applies to more scrolling devices and also removes the potential influence of selection time.

We did find that some devices do not seem to match the linear or log models (Table. \ref{tab:device_coefficients}). 
Some scrolling devices did not fit the log model well for the known condition, although they fit the linear model for unknown. This is likely because some of those techniques, like the mouse wheel with notches, the keyboard arrow keys and the scrollbar by pressing the arrow buttons have a fixed maximum speed, so users cannot go faster even if they know they have a long way to scroll. 
Other discrepancies are harder to explain. For example, the roller mouse did not fit either the predicted model in either condition: it has the lowest $R^2$ (82.63\%) for the linear model in the unknown condition, and the $R^2$ for the log model in the known condition is 82.06\%. In the known condition, the mouse wheel without notches fit the linear model ($R^2$ = 88.67\%) better than the logarithmic one ($R^2$ = 79.22\%). Additionally, the $R^2$ for two-finger scrolling on a laptop touchpad ($R^2$ = 82.28\%) and scrollbar by dragging the thumb ($R^2$ = 79.86\%) are also relatively low for the known condition, even though they do not have a fixed upper limit of speed.

With respect to finding a practical measure for accuracy (\textbf{RQ6}), the results show that the number of switchbacks is more informative than the maximum overshoot distance, because there are negative correlations in Fig. \ref{fig:accuracy_across_frame_sizes} (a) and (b) for the number of switchbacks, but not in Fig. \ref{fig:accuracy_across_frame_sizes} (c) and (d) for the max distance. Further, flicking on the iPad and iPhone have the highest values for the max overshoot distance, but are some of the quickest scrolling techniques, which further reduces our faith in using the max overshoot distance. In our observations of participants performing the user study, we found that the number of switchbacks that users had to make seemed to correlate best with our observations of how much they had to struggle and their annoyance with the scrolling techniques. So we feel that calculating the number of switchbacks is an easy and effective measure for the accuracy.

\section{Limitations and Future Work}
An obvious limitation of this paper is that we only ran \prototypeNameWithSpace on 11 people, and each of the 11 devices was only tested by 3 people, so future work should test a much larger population that is more diverse. As a partial mitigation of this issue, we run pilots on many other students in classes testing earlier versions of this design. We also did not test anyone with disabilities, and it would be interesting to measure their impact on the designs and evaluations. For example, Trewin \etal \cite{trewin2013physical} and Nicolau \etal \cite{nicolau2014mobile} attested that directional gestures (\eg flicking on touchscreen) would be challenging for people with dexterity impairment, while tapping keys or buttons would be easier and more inclusive for them. 

Despite \prototypeNameWithoutSpace's success in supporting the scrolling evaluation among different devices, there are some conditions it does not cover that would be good to support in the future. One obvious limitation is that it is currently limited to vertical scrolling only, and does not measure horizontal scrolling or panning (two-dimensional scrolling of horizontal and vertical at the same time, which can be done with the hand tool like in Adobe Photoshop or Acrobat). We think the design could be expanded to support that, potentially also adding support for measuring zooming and 3D movements.

Although we have tested \prototypeNameWithSpace on a variety of device sizes, it would currently not work on a tiny display like a smartwatch. Future work can try to shrink \prototypeNameWithSpace to run on a watch-size display like the study in \cite{chun2018qualitative}.
We also did not control for the physical size of the screen. That is, if one measured the height of the shapes with a ruler, they would clearly be smaller on a phone screen than on a laptop screen, which may affect the comparisons.

At another extreme, the current test document has a height equivalent to about 300 lines tall, which matches the size tested in previous studies and most web pages, whereas scrolling can be used for ultra-long documents. For example, it is not unknown to have 3000 contacts or 5000 songs to scroll through, or a single Microsoft Word or PDF file of a book of 500 pages. Whereas flicking is faster for all distances in our tests, clearly it will break down when the document gets to be much longer.

The current design of \prototypeNameWithSpace makes it difficult to measure the learning time for a scrolling technique, since we randomized different scrolling distances for each trial. Whereas most of our scrolling techniques were familiar to participants, some like the Roller mouse or trackball are not (Fig. \ref{fig:participant_exp}). The rate-controlled in-keyboard joystick was unfamiliar or not recently used by participants, and studies of the use of it for pointing showed significant improvements over time \cite{rutledge1990keyboard}. It would not be difficult to adapt \prototypeNameWithSpace to measure the increase in speed and accuracy over time, although participants might find it tedious to practice scrolling over and over. 

We chose a variety of current scrolling techniques to measure with \prototypeNameWithoutSpace, but there are many other scrolling techniques that could be evaluated. For example, we did not test gaze-controlled scrolling \cite{turner2015understanding} or tilt-based scrolling \cite{liu2019tilt}). We might have allowed participants to use a scrollbar in any way they wanted, rather the requiring using only the arrows or the thumb. We could also test a Macintosh scrollbar, which appears when you scroll with two-fingers and then its thumb can be dragged with the pointer. We could also test the hand tool like in Acrobat for dragging the content, which can be used for vertical scrolling like \prototypeNameWithSpace supports. The iPhone provides a scrollbar with letters for long alphabetical lists, like contacts, but this would require a change to \prototypeNameWithSpace to have targets with letters or words. 
We could also adapt \prototypeNameWithSpace so it could be used to measure auto-scroll when reaching the edge of the active area while extending the selection or performing drag-and-drop.
In addition, it would be fun to reimplement some old scrolling techniques, like those used in the original Xerox PARC Bravo editor, Smalltalk or Interlisp, or on the NeXT machine \cite{brad1990video}, and see how they compare to modern techniques.

Finally, \prototypeNameWithSpace cannot currently be used to measure ``infinite scrolling'' like on Facebook where more content is loaded when the user scrolls to the bottom, making the document larger, but it is not clear whether speed and accuracy would be useful measures in that case.

\section{Conclusion}

Having a portable and validated way to measure scrolling speed and accuracy can provide better insights into how well both new and old scrolling techniques work, and can therefore support further research into new and better scrolling mechanisms. We have developed \prototypeNameWithSpace through multiple years of iterative refinements, and believe it is now ready for general use. As shown here, it can be used across any kind of device that can run a web page, with many kinds of scrolling mechanisms. Although we emphasize scrolling without selection, an option can turn on the requirement that the user must click on the target after it is in the frame. We also provide a measure for the accuracy of the scrolling itself, which has not been analyzed by previous tests. By testing 11 different scrolling techniques, we both provide new data about the relative merits of these techniques, as well as validate that \prototypeNameWithSpace works. We hope that \prototypeNameWithSpace will be useful to the community as a way to measure and compare the speed and accuracy of their scrolling techniques, or at least, inspire more research and refinement in how such a standard measurement tool could be developed. You can use \prototypeNameWithSpace as is, or download the code and modify it from https://github.com/CharlieCRChen/scrolling-test-new/. 

\section{Acknowledgments}
We would like to thank all the participants in our user study and all the students in the Interaction Techniques class at Carnegie Mellon University in 2019 and 2022, who provided valuable feedback on the design of \prototypeNameWithoutSpace. 

\bibliographystyle{ACM-Reference-Format}
\bibliography{bibliography}

\end{document}